\documentstyle[aps,prl,preprint,floats,epsfig]{revtex}  
\begin{document}

%-- put Belle logo at top --
\topmargin -0.5in
\makeatletter
\def\maketitle{\par
\begingroup
\let\cite\@bylinecite
\def\thefootnote{\fnsymbol{footnote}}%
\if@twocolumn
\twocolumn[\@maketitle\vskip2pc]%
\else
\newpage
%\vskip -8cm
\epsfysize3cm
\epsfbox{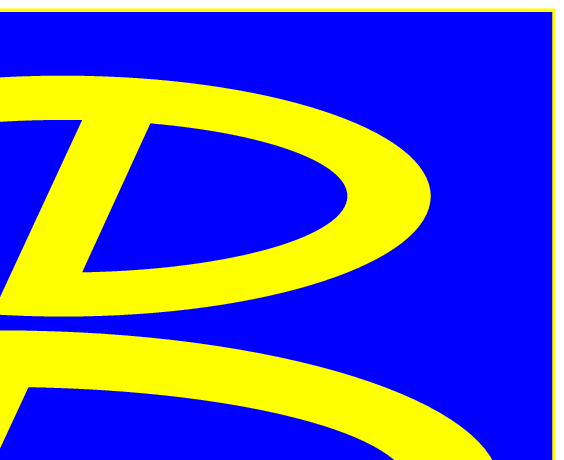}    % BELLE-logo (you need file belle.eps)
%\vskip 8cm
\global\@topnum\z@ %
\@maketitle
\fi
\thispagestyle{plain}\@thanks
\endgroup
\def\thefootnote{\arabic{footnote}}%
\setcounter{footnote}{0}%
\let\maketitle\relax \let\@maketitle\relax
\let\@thanks\relax \let\@authoraddress\relax \let\@title\relax
\let\@date\relax \let\thanks\relax
}
\makeatother

% to make it single spaced
\tighten

\draft % makes pacs numbers print
%\def\doubles{\setlength{\baselineskip}{26pt}
%             \setlength{\lineskip}{26pt}}

%\doubles

\preprint{\tighten\vbox{\hbox{\hfil KEK preprint 2000-123}
                        \hbox{\hfil Belle preprint 2000-4.v3}
                        \hbox{\hfil }
                        \hbox{\hfil }\hbox{\hfil }\hbox{\hfil }
}}
\title{
% \quad\\[1cm] \Large
 ~~\\
       Measurement of $B^0_d$-$\bar B^0_d$ mixing rate from the 
       time evolution of dilepton events at the $\Upsilon(4S)$
\footnote[0]{(to appear in Phys.  Rev. Lett.)}
}

%\author{ 
%}
%\input{belle_authors}
%\begin{center}
\author{
The Belle Collaboration  \\
(26-November, 2000; revised 1-March, 2001)
}
\author{
K.~Abe$^{8}$, 
K.~Abe$^{36}$, 
I.~Adachi$^{8}$, 
Byoung~Sup~Ahn$^{14}$, 
H.~Aihara$^{37}$, 
M.~Akatsu$^{19}$, 
G.~Alimonti$^{7}$, 
K.~Aoki$^{8}$,
K.~Asai$^{20}$, 
M.~Asai$^{9}$
% ...
}
%\end{center}

%\small
%\begin{center}
\address{
% ...
$^{7}${University of Hawaii, Honolulu HI}\\
$^{8}${High Energy Accelerator Research Organization (KEK), Tsukuba}\\
$^{9}${Hiroshima Institute of Technology, Hiroshima}\\
% ...
$^{14}${Korea University, Seoul}\\
% ...
$^{19}${Nagoya University, Nagoya}\\
$^{20}${Nara Women's University, Nara}\\
% ...
$^{36}${Tohoku University, Sendai}\\
$^{37}${University of Tokyo, Tokyo}\\
% ...
}
%\end{center}
%\setcounter{footnote}{0}
%\newpage
%\date{\today}

\maketitle

\begin{abstract}

We report  a determination of the $B^0_d$-$\bar B^0_d$ mixing parameter
$\Delta m_d$ 
based on the time evolution of dilepton yields in $\Upsilon(4S)$ decays. 
The measurement is based on a 5.9~fb$^{-1}$   
data sample collected by the Belle detector at KEKB.
The proper-time difference distributions for 
same-sign and opposite-sign dilepton events are simultaneously fitted
to an expression containing $\Delta m_d$  
as a free parameter. 
Using both muons and electrons, we obtain
$ \Delta m_d = 
0.463 \pm 0.008~({\rm stat.}) \pm 0.016~({\rm sys.})~{\rm ps}^{-1} 
  $.
This is the first determination of $\Delta m_d$  
from time evolution  measurements at the $\Upsilon(4S)$.
We also place limits on possible CPT violations.

\end{abstract}

%{\renewcommand{\thefootnote}{\fnsymbol{footnote}}

%\normalsize

% body of paper here
%\narrowtext
%\twocolumn

The frequency of $B^0_d$-$\bar B^0_d$ mixing is 
proportional to the mass difference between the two 
mass eigenstates of the neutral $B$ meson, $\Delta m_d$,
and is a fundamental parameter of  the $B$  system. 
Measurements of $\Delta m_d$ derived from
the time evolution of $B^0_d$ decays have been reported 
by CDF, SLD, and the LEP experiments \cite{LEPCDFSLDmix}; 
ARGUS and CLEO have
measured it using the integrated fraction of same-flavor 
$B$ pair decays in $\Upsilon(4S)$ events \cite{ARGUSmix,CLEOmix}. 
We report here the first determination of 
$\Delta m_d$ 
based on the time evolution of $B^0_d$ decays
in $\Upsilon(4S)$ events produced in 
asymmetric  $e^+e^-$ collisions, using data collected by the Belle detector
\cite{BELLE}
at the KEKB storage ring\cite{KEKB}.

At the $\Upsilon(4S)$, the asymmetry in
time evolution between same-flavor ($B^0_d B^0_d$, $\bar B^0_d \bar B^0_d$) 
and opposite-flavor ($B^0_d \bar B^0_d$) decay pairs
exhibits an oscillation as a function of the proper time 
difference between the two $B$-meson decays,  $\Delta t$, with a frequency
that is proportional to $\Delta m_d$.
In KEKB, collisions between 8.0~GeV  electrons and 3.5~GeV positrons
have a center of mass (CM) motion along the electron 
beam direction (z direction) with a Lorentz boost of $\gamma \beta$ = 0.425.  
Since each of the two $B$'s 
is produced nearly at rest in the CM,
the separation
of their decay vertices in the lab frame is proportional to $\Delta t$
and has an average magnitude of 200$\mu$m.
High-momentum leptons can be used both for tagging the $B$ flavor
and for determining the decay vertex with good accuracy. 
The $\Delta t$
in dilepton events can thus be used to measure 
the time evolution of $B$ decays. 
The same analysis can be used to test CPT conservation by 
the inclusion of the 
complex parameter 
$\cos\theta$ in the fit\cite{CPT}.

The analysis presented here is based on  integrated 
luminosities of 5.9~fb$^{-1}$ 
at the $\Upsilon(4S)$ resonance and 0.6~fb$^{-1}$ at an energy
that is 60 MeV below the peak.

The Belle detector  
consists of a silicon vertex detector (SVD),
a central drift chamber (CDC), 
an array of 1188 aerogel Cerenkov counters (ACC), 
128 time-of-flight (TOF) scintillation counters, 
and an electromagnetic calorimeter containing 8736 CsI(Tl) 
crystals (ECL), 
all located 
inside the 3.4-m-diameter superconducting solenoid that generates
a 1.5~T magnetic field.
An iron return yoke, outside the solenoid,
is segmented into 14 layers of 4.7-cm-thick iron 
plates alternating with a system of resistive plate counters that is  
used to identify muons and $K_L$ mesons (KLM).

Hadronic events are required to have at least five tracks,
an event vertex with radial and $z$ coordinates respectively
within 1.5 cm and 3.5 cm of the origin,
a total reconstructed CM energy greater than 0.5$W$
($W$ is the $\Upsilon(4S)$ CM energy),
a $z$-component of the net reconstructed CM momentum less than 
$0.3W/c$,  
a total CM calorimeter energy between $0.025W$ and $0.90W$, and
a ratio $R_2$ of the second and zeroth Fox-Wolfram 
moments\cite{foxwolfram} that is less than 0.7.
While the $R_2$ cut suppresses events of non-$\Upsilon (4S)$ origin, 
all other cuts are intended to remove the beam-related background and
QED events. 

For electron identification, we use position, 
cluster energy, 
and shower shape in the ECL, $dE/dx$ 
in the CDC, and hit information in the ACC.
This is $\sim 90$\% efficient for electrons and has a
$\sim 0.3$\% misidentification probability for charged hadrons 
with momenta above 1 GeV/$c$.
Electrons from $\gamma$ conversions are removed.

Muon selection is based on KLM hits associated with
charged tracks.
The range of the tracks and the
matching quality of the hits are used.
The efficiency is $\sim 85\%$ for muons with
momentum above 1~GeV/$c$
and the misidentification probability is $\sim 2\%$.

Events containing leptons from $J/\psi$ decays are rejected.
In addition, lepton candidates are required 
to satisfy
$ 30^\circ < \theta < 135^\circ$;
      $1.1~{\rm GeV}/c < p^* < 2.3~{\rm GeV}/c$;
      $\mid dr^{\rm IP} \mid < 0.05$~cm;
      $\mid dz^{\rm IP} \mid <2.0$~cm;
and have at least one (two) associated SVD hit(s) 
     in the $r$-$\phi$ ($r$-$z$) view,
where $\theta$ is the laboratory polar angle,
$p^*$ is the CM momentum, and
$dr^{\rm IP}$ and $dz^{\rm IP}$ are the distances of closest
      approach to the run-dependent interaction point.
To reduce secondary leptons and fakes from the 
same $B$ and from the continuum,
which tend to be back-to-back,
the opening angle $\theta_{\ell\ell}^*$ between the leptons in 
the CM frame is required to satisfy 
$-0.8 < \cos \theta_{\ell\ell}^* < 0.95$.
The application of the above-listed criteria yields 
8573 same-sign (SS) and 40981 opposite-sign (OS) dilepton events
on the $\Upsilon(4S)$, and 40 SS and 198 OS dilepton events
below the resonance.

The $z$-vertex of
leptons is determined from the
intersection of the  
lepton tracks with the profile of
$B^0_d$ decay vertices,  
 which is estimated from the profile of the beam interaction point (IP) 
convolved with the average $B$ flight length
    ($\sim 20\mu$m in the $\Upsilon(4S)$ rest frame).
The mean position and the width
    $(\sigma_x^{IP},\sigma_y^{IP},\sigma_z^{IP})$ of the IP are
     determined on a run-by-run basis  using hadronic events.
We find $\sigma_x^{IP}=100$-$120\mu$m, $\sigma_y^{IP}\sim 5\mu$m and
     $\sigma_z^{IP}=2$-$3$mm.
The proper-time difference is calculated from the $z$ positions of 
the two lepton vertices using the relation
 $  \Delta t ~=~ {\Delta z}/{c\beta\gamma}, $
where $\Delta z = z_1 - z_2$ is the difference between the two $z$ 
vertices.
For OS events, the  
positively charged lepton is taken as the first lepton ($z_1$).
For SS
events the absolute value of $\Delta z$ is used.

The observed SS and OS dilepton proper-time distributions
have contributions from ``signal,'' defined 
as events where both leptons are primary 
leptons from semileptonic decay of $B^0_d$ or $B^+$, and
``background,'' where
at least one lepton is secondary or fake, or the event is 
from the non-$\Upsilon(4S)$ continuum.
The value of $\Delta m_d$ was extracted by 
simultaneously fitting the two distributions to
the respective
sums of contributions from all known signal and background sources.

Each dilepton is identified with one of the event types listed in 
Table~\ref{contrib}.
Each event type is categorized as either
signal (S), correctly tagged background (C), or
incorrectly tagged background (W).
For each, we  parameterize the proper-time distribution
as the product of the number
of contributing events ($N$), an overall selection efficiency ($\epsilon$),
and a normalized distribution function $\tilde P(\Delta t)$.
The values of $N$ depend on the total number of
$\Upsilon$(4S) events in
the data sample ($N_{4S}$), the
branching fractions of the $\Upsilon$(4S) to neutral and charged $B$ pairs
  ($f_0$ and $f_+=1-f_0$),
the semileptonic branching fractions
($b_0$ and $b_+$ for neutral and charged $B$),
and, for neutral $B$'s, the mixed event fraction
$\chi_d = x_d^2/(2(1 + x_d^2))$
  where      $x_d = \tau_{B^0_d}\Delta m_d $
and $\tau_{B_d^0}$ is the $B_d^0$ lifetime.
The evaluations of $N$ are summarized in
Table~\ref{contrib}.
The efficiencies $\epsilon$ are determined by Monte Carlo (MC) simulation.

The observed proper-time distribution function $\tilde P(\Delta t)$ 
for the signal is a
convolution of a root distribution with a detector response 
function, $g$,
and is given by:
\begin{equation}
 \tilde P(\Delta t) =  {\int g( \Delta t - \Delta t^{\prime}) 
       F(\Delta t^\prime) d(\Delta t^\prime)
\over \int \int g( \Delta t - \Delta t^{\prime}) 
       F(\Delta t^\prime) d(\Delta t^\prime)d(\Delta t)},           
\end{equation}
where 
the respective theoretical root
functions for mixed $B^0$, unmixed $B^0$,
and charged $B$
 are
\begin{eqnarray} 
F(\Delta t) &=&
       (1 / 4\tau_{B^0_d} ) 
      e^{-|\Delta t|/\tau_{B^0_d}} [1 - 
\cos(\Delta m_d \Delta t)] \label{eq:theorB0SS} \\
 F(\Delta t) &=&
      (1 / 4\tau_{B^0_d} ) 
      e^{-|\Delta t|/\tau_{B^0_d}} [1 + \cos(\Delta m_d \Delta t)]
\label{eq:theorB0OS}  \\
 F( \Delta t) &=&
      (1 / 2\tau_{B^+}) e^{-|\Delta t|/\tau_{B^+}}.\label{eq:theorBPOS}
\end{eqnarray}

MC simulations of generic $B^+B^-$, unmixed and mixed $B^0$,
and continuum events are used 
to determine the background $\Delta z$ distributions.
The dominant background source is a primary lepton paired with a 
secondary lepton from a $c$-quark.
The shape as well as the normalization of the background from
neutral $B$ events depends on $\Delta m_d$.
To account for this, we generated two samples of generic 
neutral $B$ events, one with
$\Delta m_d=0.464~$ps$^{-1}$ and one with $\Delta m_d=0.423~$ps$^{-1}$.
Background distributions for arbitrary $\Delta m_d$ are
determined by linear interpolation.

We use the $\Delta z$ distribution of dileptons from  $J/\psi$ decays 
in the data for $g$;  
for these events the root 
distribution is  a delta function
and the lepton momentum spectra are in the same region as those of 
primary leptons from $B$ decays.
The $\Delta z$ distribution for $J/\psi$ events, which has
$\sigma=112~\mu$m,  agrees with the 
MC distribution if it is 
convolved with a Gaussian of $\sigma=50\pm 18~\mu$m. 
This is due to an imperfect detector simulation, and we correct this
effect by applying a convolution with $\sigma=50~\mu$m to each 
MC-determined background distribution.

\begin{figure}[!htb]
\centerline{
\epsfxsize 6.0 truein \epsfbox{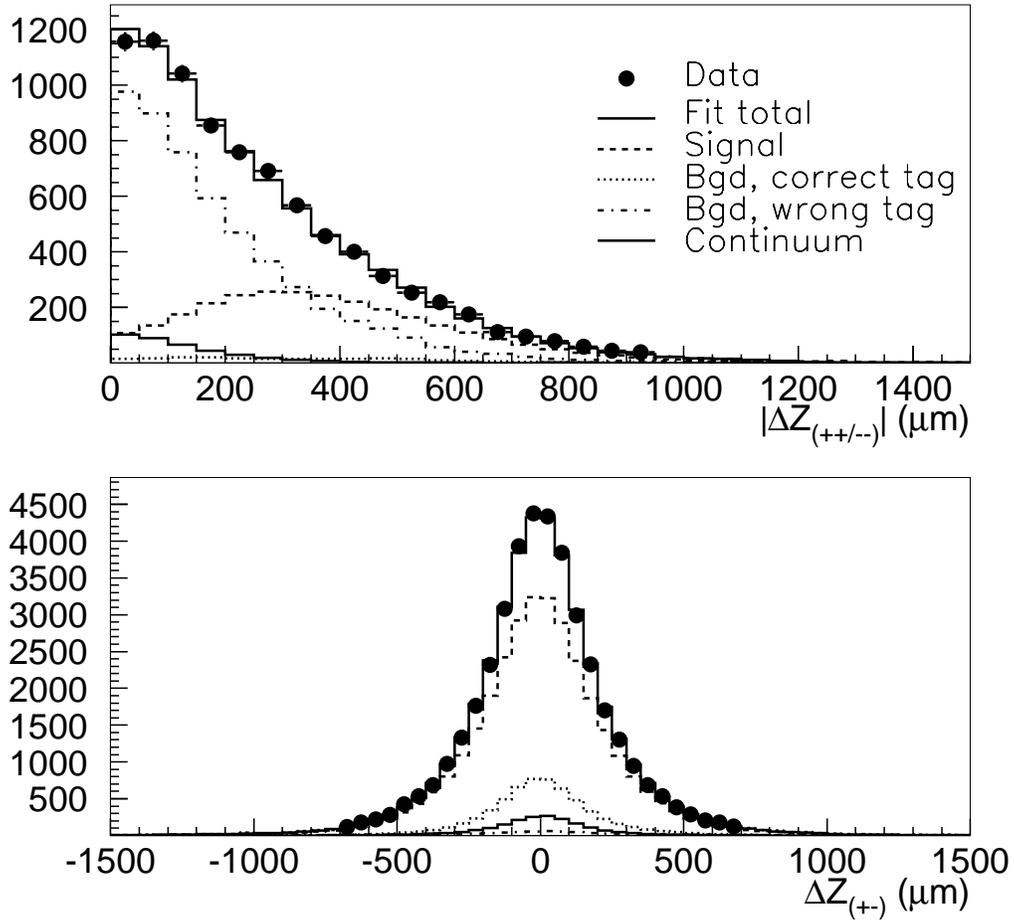}
}   
\caption{ $\Delta z$ distribution of dileptons for data together with
          fit result. The upper plot shows the distributions for same-sign, 
  and the lower plot for opposite-sign dileptons.
  Signal and background dileptons obtained from the fit are also shown. }
\label{dl_fit}
\end{figure}

\begin{figure}[!htb]
\centerline{
\epsfxsize 6.0 truein \epsfbox{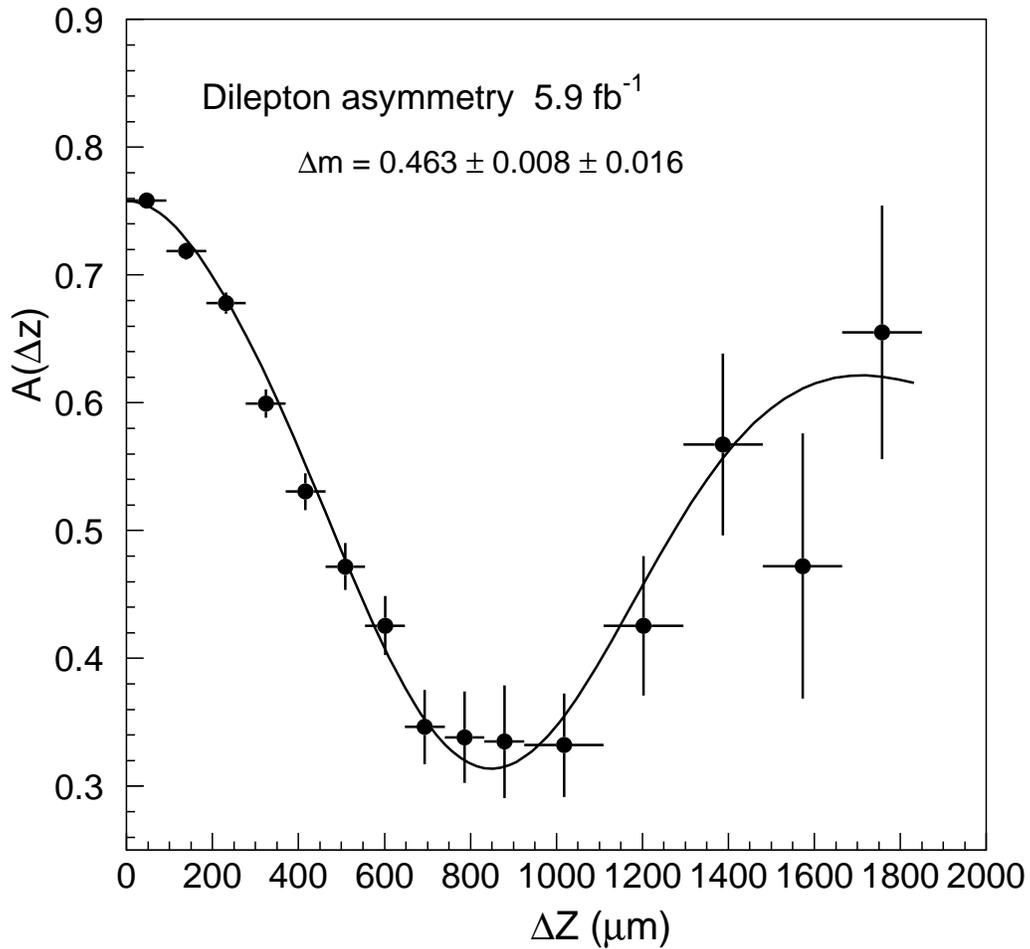}
}   
\caption{ Opposite and same-sign dilepton asymmetry vs $\Delta z$.
The asymmetry is defined as 
$A(\Delta z)$ = $(N_{\rm OS} - N_{\rm SS})/(N_{\rm OS} + N_{\rm SS})$.
The points are the data.  
The curve is the result of the fit.
}
\label{dl_asym}
\end{figure}

To extract $\Delta m_d$, a binned maximum likelihood fit 
is performed simultaneously to the $\Delta z$
distributions of the SS and OS dileptons.
Each fitting function is a sum of signal and background
distributions.
In order to properly take into account the tails of the $\Delta z$
distributions, the signal response function and the background
distribution are given in the form of a lookup table rather than
an analytic function.
We fix the parameters $\tau_{B^0_d}=1.548$ ps\cite{PDG},
$f_+/f_0 = 1.05 $\cite{CLEO_f0}, 
and $\tau_{B^+}/\tau_{B^0_d} = 1.06$\cite{PDG}, 
and limit the fit region to $|\Delta z| <$ 1.85~mm.  
The constraint $b_+/b_0 = \tau_{B^+}/\tau_{B^0_d}$ is imposed.
The continuum contribution 
is fixed to that of off-resonance data, scaled to account for luminosity
and energy differences.
The relative selection efficiencies for the event types (mixed
$B_d^0$, unmixed $B_d^0$, and charged $B$) within each tag type 
(S, C, and W) are fixed, resulting in
two free parameters (efficiency ratios in C/S and W/S) in addition to 
$\Delta m_d$ and the overall normalization.
The fit result is
$ \Delta m_d = 0.463 \pm 0.008~{\rm ps}^{-1}$  
with $\chi^2/DOF = 333/376$.
The efficiency ratios in C/S and W/S are 
$(9.66 \pm 1.39)\times 10^{-3}$ and
$(6.98\pm 0.25)\times 10^{-3}$,
respectively, 
which give signal fractions to be 32.1\%(SS) and 77.5\%(OS).
Figure~\ref{dl_fit} shows the $\Delta z$ distributions for the data  
together with the fitted curves. 
Fig.~\ref{dl_asym} shows the OS and SS asymmetry,
$(N_{\rm OS}-N_{\rm SS})/(N_{\rm OS}+N_{\rm SS})$, for data together with 
the result of the fit.

As a cross check, we also measured $\Delta m_d$ using a fitting method
that differs from the one described above in the following 
aspects\cite{TOMOTO}:
a) an unbinned rather than binned maximum likelihood fit;
b) response function is the sum of three Gaussians, with parameters
   determined 
   from the dileptons from $J/\psi$ decays;
c) backgrounds separated into SS, OS, rather than C, W;
d) background distributions were analytic functions, with parameters
   determined by fitting to MC
We find
$ \Delta m_d$ in the range 0.460 to 0.483 ${\rm ps}^{-1} $
depending on the choice of analytic forms for the backgrounds,
which is consistent with the primary result.

The systematic errors were estimated by repeating the
fits for different input parameters.
The main contribution
originates
from uncertainties on input parameters and from 
determination of the response functions.
Contributions 
of $f_+/f_0$, $\tau_{B^0_d}$ 
and $\tau_{B^+}/\tau_{B^0_d}$ are estimated by 
adjusting each in turn
by the amount of its uncertainty.  
Contribution of the response function 
arises from the possibility that
it differs from the true dilepton response function,
from the statistical uncertainties of the determination, and
from the fact that the calculation of proper time 
$ \Delta t ~=~ \Delta z/c\beta\gamma $
is not exact due to
the motion of the $B$'s in the CM and the energy spread of the beams. 
To estimate the first possibility, we 
used the
MC dilepton response function convolved
with a Gaussian of $\sigma =50\mu$m.
For the second, we 
varied the 
number of entries on a bin-by-bin basis by the amount of the
statistical errors. 
For the third, we
compared two fits, one using a  response function obtained 
for the true $\Delta t$ difference and a second obtained 
for $\Delta z$.

We also consider
the uncertainty from the background simulation. 
We assigned a $35\%$ error for the fake rate and adjusted the fake rate
by this amount.
We varied the branching ratios 
of $B$ decaying
to $D^0$ and $D^+$
in the MC
in accord with the experimental uncertainties\cite{CLEO_sec}.
We varied the width of the Gaussian used to correct for an
imperfect detector simulation by $\pm 18 \mu$m.

It is assumed in the fit that $\Delta\Gamma$, the difference between
the decay widths of the neutral $B$ mass eigenstates, is zero.
Although no significant experimental constraint exists\cite{CLEOmix},
it is predicted based on solid theoretical grounds to be
very small ($\Delta\Gamma/\Gamma<$ 1\%)\cite{PDG}.
We repeated the fit including the effects of $\Delta\Gamma/\Gamma$ = 
1\% and found the shift in the result to be negligible ($<$0.001 ps$^{-1}$).

Contributions to the systematic error from the above sources 
are summarized in Table~\ref{syserr}.
The total systematic error is obtained by summing all errors
in quadrature:
$$ \Delta m_d = 
   0.463 \pm 0.008~({\rm stat}) \pm 0.016~({\rm sys})~{\rm ps}^{-1}.  
$$

When the constraint of CPT conservation is removed in 
$B^0_d$-$\bar B^0_d$ mixing, the theoretical functions 
(\ref{eq:theorB0SS}) and (\ref{eq:theorB0OS}) are modified and
become
\begin{eqnarray}
 F(\Delta t) &=&
      (|\sin\theta|^2 / 4\tau_{B^0_d} ) 
      e^{-|\Delta t|/\tau_{B^0_d}} [1 - \cos(\Delta m_d \Delta t)]\\
 F(\Delta t) &=&
      (1 / 4\tau_{B^0_d} ) 
      e^{-|\Delta t|/\tau_{B^0_d}} [1 + |\cos\theta|^2
\nonumber \\ & &
      + (1 - |\cos\theta|^2)\cos(\Delta m_d \Delta t)
\nonumber \\ & &
      - 2 Im(\cos\theta)\sin(\Delta m_d \Delta t) ] 
\end{eqnarray}
and $\chi_d$ becomes
$ (|\sin\theta|^2 x_d^2)  /
          [ |\sin\theta|^2 x_d^2 + 
            (2 + x_d^2 + x_d^2 |\cos\theta|^2) ] $.
A non-zero value of the complex parameter $\cos\theta$
would be an indication of CPT violation.
The result of the fit is\cite{LEO}
\begin{eqnarray}
  Im(\cos\theta) &=& 0.035 \pm 0.029 {\rm (stat)}   
                      \pm 0.051 {\rm (sys)}
\nonumber \\
 Re(\cos\theta) &=& 0.00  \pm 0.15  {\rm (stat)} \pm 0.06 {\rm (sys)} 
\nonumber
\end{eqnarray}
and $\Delta m_d$ = 0.461 ps$^{-1}$.
These results imply\cite{CPT}
the upper limits
$|{m_{B^0}-m_{\bar B^0}| / m_{B^0}}  < 1.6 \times 10^{-14}$
and
$ |{\Gamma_{B^0} - \Gamma_{\bar B^0}| / \Gamma_{B^0}}  < 0.161 $
at the 90\% C.L.

In summary, we report
the first determination of $\Delta m_d$ 
using the time evolution of $B^{0}$-mesons produced in 
$\Upsilon(4S)$ decays. 
We obtain
$ \Delta m_d = 
   0.463 \pm 0.008~({\rm stat}) \pm 0.016~({\rm sys})~{\rm ps}^{-1} 
$,
which is consistent with the world average value
$\Delta m_d = 0.472 \pm 0.017 ~{\rm ps}^{-1}$\cite{PDG}. 
We have also examined CPT violation and obtain
the first limit on  
$ ({m_{B^0}-m_{\bar B^0})/ m_{B^0}}$
and a limit on      
$ (\Gamma_{B^0} - \Gamma_{\bar B^0})/ \Gamma_{B^0}$
that is compatible with the previous
measurement\cite{IMS}.

%\section*{Acknowledgments}
We wish to thank the KEKB accelerator group 
for the excellent operation.
We 
acknowledge support from the Ministry of Education, Science, Sports and
Culture of Japan and
the Japan Society for the Promotion of Science;
the Australian Research Council and the Australian Department of Industry,
Science and Resources;
the Department of Science and Technology of India;
the BK21 program of the Ministry of Education of Korea and
the Basic Science program of the Korea Science and Engineering Foundation;
the Polish State Committee for Scientific Research 
under contract No.2P03B 17017; 
the Ministry of Science and Technology of Russian Federation;
the National Science Council and the Ministry of Education of Taiwan;
the Japan-Taiwan Cooperative Program of the Interchange Association;
and  the U.S. Department of Energy.

\newpage
\begin{table}[!thb]
\begin{center}
\caption {Categorization of event types contributiong to dilepton events.}
\label{contrib}
%\vspace{0.3cm}
\begin{tabular}{l l l c}
\multicolumn{2}{c}{$\ell\ell$  event type} & $N$ & tag type\\
\hline
SS signal & $B^0_d$, mixed &   $N_{4S} f_0 \chi_d b_0^2$ & S \\ 
\hline
SS background& $B^0_d$, mixed &      $ N_{4S} f_0\chi_d$ &C \\
             & $B^0_d$, unmixed   &  $ N_{4S}f_0 (1-\chi_d)$ &W\\
             & $B^+ B^-$  & $  N_{4S} f_+$ &W\\
             & continuum & $ N_{\rm cont}$ & \\
\hline
OS signal    & $B^0_d$, unmixed &  $ N_{4S} f_0(1-\chi_d) b_0^2$  &S\\
             & $B^+B^-$&   $ N_{4S} f_+ b_+^2$ &S\\
\hline
OS background& $B^0_d$, mixed & $ N_{4S} f_0 \chi_d $ &W \\
             & $B^0_d$, unmixed  & $  N_{4S} f_0 (1-\chi_d)$ &C \\
             & $B^+ B^-$ & $  N_{4S} f_+ $ &C \\
             & continuum & $  N_{\rm cont}$ & 
\end{tabular}
\end{center}
\end{table}

\begin{table}[!thb]
\begin{center}
\caption { Summary of systematic errors.}
\label{syserr}
\vspace{0.3cm}
\begin{tabular}{l c }
Source (uncertainty)   & $\Delta m_d$  
\\ \hline \hline
$f_+/f_0$ ($\pm$ 0.08) & $\pm$0.009 
 \\
$B^0_d$ lifetime ($\pm$ 0.032~ps) & $\pm 0.004$ 
\\
$\tau_{B^+}/\tau_{B^0_d}$ ($\pm$ 0.03) & $\pm 0.009$ 
\\
response function & $\pm$0.005
\\
background fake rate ($\pm35\%$) &   $\pm 0.004$ 
 \\
${\cal B}(B \rightarrow D X)$  ($D^0$: $\pm 4.6\%$, $D^+$: $\pm 14.3\%$)
             & $ \pm 0.002 $ 
\\
continuum  (SS: $\pm 16\%$, OS: $\pm 7\%$) & $\pm 0.002$ 
\\
detector resolution, BG ($\pm 18~\mu$m) & $\pm 0.001$ 
\\
Monte Carlo statistics  & $\pm 0.004$ 
\\ \hline
total  & $\pm 0.016$ 
\end{tabular}
\end{center}
\end{table}


\end{document}